\documentclass[12pt]{article}
\usepackage{amsmath, amsfonts, amssymb}
\usepackage{color}
\usepackage{psfrag}
\usepackage{graphicx}
\usepackage{subfigure}
\usepackage{rotating}
\usepackage{yfonts}

\newcommand{\bea}{\begin{eqnarray}}
\newcommand{\eea}{\end{eqnarray}}
\newcommand{\be}{\begin{equation}}
\newcommand{\ee}{\end{equation}}
\newcommand{\vs}[1]{\vspace{#1 mm}}

\renewcommand{\a}{\alpha}
\renewcommand{\b}{\beta}
\newcommand{\g}{\gamma}
\newcommand{\de}{\delta}
\newcommand{\e}{\epsilon}

\newcommand{\La}{\Lambda}
\newcommand{\ka}{\kappa}
\newcommand{\om}{\omega}
\newcommand{\Om}{\Omega}
\newcommand{\half}{\frac{1}{2}}
\newcommand{\pa}{\partial}
\newcommand{\s}{\sigma}
\newcommand{\Si}{\Sigma}

\newcommand{\nn}{\nonumber}

\newcommand{\cg}{{\mathfrak{g}}}
\newcommand{\gca}{{\mathfrak{gca}(2)}}

\begin{document}
\topmargin 0pt
\oddsidemargin 0mm

\begin{flushright}



\end{flushright}

\vspace{2mm}

\begin{center}
{\Large \bf Wess-Zumino-Witten Model for Galilean Conformal Algebra}

\vs{10}

{
%

Somdeb Chakraborty\footnote{E-mail: somdeb.chakraborty@saha.ac.in},
Parijat Dey\footnote{E-mail: parijat.dey@saha.ac.in}}


{\em
 Theory Division, Saha Institute of Nuclear Physics,\\
 1/AF Bidhannagar, Kolkata-700 064, India\\}

\end{center}

\vs{10}

\begin{abstract}
In this note, we construct a Wess-Zumino-Witten model based on the Galilean conformal algebra in $2$-spacetime dimensions, which is a nonrelativistic analogue of the relativistic conformal algebra. We obtain exact background corresponding to $\s$-models in six dimensions (the dimension of the group manifold) and a central charge $c=6$. We carry out a Sugawara type construction to verify the conformal invariance of the model. Further, we discuss the feasibility of the background obtained as a physical spacetime metric.
\end{abstract}

\newpage


The Wess-Zumino-Witten (WZW) model is a typical example of a conformal field theory in $2$-dimensions. They have been studied extensively in the context of string theory since they are known to yield a class of exact string backgrounds. Given a semisimple Lie algebra $\cg$, it is fairly straight forward to construct a WZW model out of it. But when one attempts to carry out a similar construction based on nonsemisimple algebras one encounters complications.
The WZW action on a Riemann surface $\Si$ is defined through
\be   \label{WZWact}
S_{WZW}(g)=\frac{1}{4\pi}\int \limits_{\Si}d^{2}\s \Om_{AB}A^{A}_{\a}A^{B \a}+ \frac{i}{12 \pi}\int \limits_{B, \pa B \equiv \Si}d^{3}\s \e_{\a \b \g}\Om_{CD}f^{D}_{~AB}A^{A \a}A^{B \b} A^{C \g}
\ee
(a sum over repeated indices is implicit) where the indices $\a, \b, \g$ are defined on the $3$-manifold $B$ whereas the indices $A,B,C,D$ are defined on the group manifold $G$ with algebra $\cg$ and generators $T^{A}$ and the gauge fields $A^{A}_{\a}$'s are defined through $g^{-1}\pa_{\a}g=A^{A}_{\a}T_{A}$.  $B$ is a $3$-manifold bounded by $\Si$, i.e., $\pa B = \Si$. The fields are mappings from the Riemann surface $\Si$ to the target manifold $G$, which is equipped with a metric structure $\Om_{AB}$. Further, we have extended the map $g: \Si \rightarrow G$ to $g: B \rightarrow G$ in an unspecified way. As is evident above, the construction of the WZW action from a Lie algebra $\cg$ requires the existence of a bilinear form in the generators $T_{A}$ and a natural choice is to take this to be the Cartan-Killing form \footnote{We shall use $\Om_{AB}$ to denote the nondegenerate bilinear form. Later, we shall use $\om_{AB}$ to denote  the degenerate Cartan-Killing metric defined in terms of the structure constants.}
\be 
\Om_{AB}=f^{~~D}_{AC}f^{~~C}_{BD} 
\ee
where the structure constants are defined as $[T_{A},T_{B}]=f^{~~C}_{AB}T_{C}$. One of the hallmarks of nonsemisimple algebras is the fact that the Cartan-Killing form becomes degenerate (noninvertible) which renders it unsuitable as a well-behaved metric on the group manifold $G$. One is then compelled to explore alternative avenues to find a suitable bilinear form which is well-defined even for nonsemisimple algebras and hence, can be used for the WZW construction. It was first shown by Nappi and Witten \cite{NW} how to tackle this problem for the case of $2$-dimensional Euclidean group with central extension and they obtained a $\s$-model describing string propagation in $4$-dimensional gravitational  plane wave background. The analysis was subsequently generalized to $d$-dimensional centrally extended Euclidean algebra in \cite{Sfe}. Further work in this direction can be found in \cite{OliRab, Sfe1, Sfe2, Moh, KehMee, Keh}. 
In this note we shall construct a WZW model based on the Galilean conformal algebra (GCA) in $2$-dimensions which we denote as $\gca$. GCA is a nonsemisimple algebra and physically it is the algebra of nonrelativistic conformal symmetries. This makes it significant in the study of condensed matter systems, which are essentially nonrelativistic. Recently, the AdS/CFT correspondence has emerged as a key tool for the investigation of such condensed matter systems near phase transition by establishing their duality with suitable gravity systems. In this context, GCA has the potential to play a vital role in our endeavour to generalize the AdS/CFT correspondence to a nonrelativistic setting and study various real life strongly interacting systems. This provides the primary motivation for studying the GCA in more detail. In this note we have attempted to construct a WZW model starting from $\gca$, driven by the curiosity whether it can yield an exact string 
 background. Along the way we also find a new nondegenerate quadratic form defined over the group manifold by alluding to an isomorphism between $\gca$ and the Poincar\'e algebra in $(2+1)$-dimensions,  $iso(2,1)$. 


Before proceeding further, let us briefly review the essential features of the GCA. Conformal Galilean algebra (and the closely related Schr\"{o}dinger group) has been studied extensively in various contexts for a long time (see, for example, \cite{Baru, HavPleb, HenUnt, LukStich, DuvHor, DuvHor2}). Recently, it attracted renewed interest when Bagchi and Gopakumar  \cite{BagGop} tried to exploit it in their attempts to systematically arrive at a nonrelativistic version of the AdS/CFT correspondence. Later it was discussed in the context of $2$-dimensional spacetime by Bagchi \textit{et al} in \cite{BagGop1}. GCA is obtained by the well-known method of group contraction applied to the relativistic conformal group $SO(d+1,2)$ in $(d+1)$-spacetime dimensions in the same way as the Galilean algebra is obtained from the Poincar\'e algebra.
The generators of GCA obey the algebra
\bea
\begin{aligned}
&[L^{(m)},L^{(n)}]=(m-n)L^{(m+n)},
\\
& [J_{ij},M^{(m)}_{k}]=-\left(M_{i}^{(m)}\de_{jk}-M_{j}^{(m)}\de_{ik}\right),
\\
&[L^{(m)},M_{i}^{(n)}]=(m-n)M_{i}^{(m+n)}
\end{aligned}
\eea
along with the standard commutation between $[J_{ij},J_{kl}]$ and the remaining commutators vanish. Here $m,n=0,\pm 1$ and $i,j,k$ denote spatial indices. One can see that the generators $L^{(m)}$ span a $sl(2,R)$ subalgebra. Another remarkable result is that the finite-dimensional GCA, as given above, can be extended to an infinite-dimensional algebra (also sometimes referred to as GCA) by uplifting $n$ to an \textit{arbitrary} integer and redefining $J_{ij}$ as $ J^{(n)}_{ij}= -t^{n}(x_{i}\pa_{j}-x_{j}\pa_{i})$.\\
In this note we shall consider $\gca$ with $n=0,\pm 1$ which follow the algebra
\begin{subequations}  \label{GCA2d}
\be \label{lsub}
[L^{(m)},L^{(n)}]=(m-n)L^{(m+n)},
\ee
\be 
[L^{(m)},M^{(n)}]=(m-n)M^{(m+n)},
\ee
\be  \label{msub}
[M^{(m)},M^{(n)}]=0.
\ee
\end{subequations}
Henceforth, we shall denote the generators by $\{ L^{(0,\pm)},M^{(0,\pm)}\}$. The Abelian subalgebra in Eq.(\ref{msub}) is responsible for making $\gca$ nonsemisimple. The degenerate Cartan-Killing form is given by \cite{BagKun},
\be  \label{CKform}
\om_{AB}=f^{~~D}_{AC}f^{~~C}_{BD}=\begin{pmatrix}
                                0& 0 &  -2  & 0 & 0 & 0\\
  0 & 1 & 0  & 0 & 0 & 0 \\
  -2 & 0 & 0  & 0 & 0 & 0 \\
  0 & 0 & 0 & 0 & 0 & 0  \\
 0 & 0 & 0 & 0 & 0 & 0  \\
 0 & 0 & 0 & 0 & 0 & 0 
                              \end{pmatrix}
\ee
where the indices $A,B$ run from $1$-$6$ and we have performed the following identification between the indices of the generators of $\gca$ and the generators $T_{A}$ of a Lie group.
\be    \label{identification}
\left(L^{(-)},L^{(0)},L^{(+)},M^{(-)},M^{(0)},M^{(+)}\right) \longmapsto \left(T_{1},T_{2},T_{3},T_{4},T_{5},T_{6}\right)
\ee
for computing $\om_{AB}$. The upper left $3 \times 3$ nondegenerate block stems from the $sl(2,R)$ subalgebra spanned by $L^{(0,\pm )}$. However, $\gca$ does allow for a nondegenerate $2$-form defined over the whole group manifold, similar to the Nappi-Witten algebra \cite{NW} or the Abelian extension of the $d$-dimensional Euclidean algebra \cite{Sfe}. The general way to construct a nondegenerate $2$-form for such nonsemisimple algebra goes by the  name of double extension as discussed in \cite{FigSta}. However, here we shall obtain the nondegenerate bilinear form in a different way by exploiting an isomorphism between $iso(2,1)$ and $\gca$, first discussed in the context of the BMS-GCA correspondence in \cite{BagPRL}.\\
The  quadratic form $\Om_{AB}$ that we are seeking has to respect the following constraints:
\bea\begin{aligned} \label{cons}
& \Om_{AB}=\Om_{BA},\\
& f^{~~D}_{AB}\Om_{CD}+f^{~~D}_{AC}\Om_{BD}=0,\\
& \Om_{AB}\Om^{BC}=\de^{~C}_{A}.
\end{aligned}
\eea
It turns out that the algebra given in Eq.(\ref{GCA2d}) is isomorphic to the $(2+1)$-dimensional Poincar\'e algebra $iso(2,1)$,
\be
[J_a,J_b] = \epsilon_{abc} J^c, \quad [J_a,P_b]~=~\epsilon_{abc}P^c, \quad  [P_a,P_b]~=~0
\ee
where the indices $a,b,c$ take values $1,2,3$. $ISO(2,1)$ is a $6$-dimensional nonsemisimple group and apart from the usual quadratic Casimir $C^{(1)}=P_{a}P^{a}$, admits an additional one - the helicity $J_{a}P^{a}$,\footnote{$iso(2,1)$ can be obtained as a contraction of $sl(2,R) \times sl(2,R)$, the isometry group of $AdS_{3}$, which admits two bilinear forms. However, only one of them remains nondegenerate in the flat space limit. This is the bilinear form we use here.}
\be
C^{(2)}=J_{a}P^{a}=\tilde{\ka}^{AB}\tilde{T}_{A}\tilde{T}_{B}
\ee
with $\tilde{T}_{A}$ being the generators of $ISO(2,1)$ with the identification
\be \nn
\{J_{1},J_{2},J_{3},P_{1},P_{2},P_{3} \} \longmapsto \{\tilde{T}_{1},\tilde{T}_{2},\tilde{T}_{3},\tilde{T}_{4},\tilde{T}_{5},\tilde{T}_{6} \}.
\ee
Upon performing the following identification
\begin{subequations}
\be 
J_1= \frac{1}{2}\left(L^{(+)} + L^{(-)}\right),  \quad J_2 = \frac{1}{2i}\left(L^{(-)} - L^{(+)}\right),  \quad J_3 = -i L^{(0)},
\ee
\be
P_1 = \frac{1}{2}\left(M^{(+)} + M^{(-)}\right),  \quad P_2 = \frac{1}{2i}\left(M^{(-)} - M^{(+)}\right),  \quad P_3 = -i M^{(0)}
\ee
\end{subequations}
we can easily establish the isomorphism between $iso(2,1)$ and $ \gca$. The nondegenerate quadratic form corresponding to $C^{(2)}$ in the basis $\{J_{a},P_{a}\}$ is
\be
\tilde{\ka}^{AB} = \half  \begin{pmatrix}
  0 & 0 &  0 &  1 & 0 & 0  \\
  0 & 0 &   0 & 0 & 1& 0  \\
  0 & 0 & 0 & 0 & 0 & 1 \\
  1 &  0 &  0 &  0  & 0 & 0 \\
  0 &  1 & 0 & 0 & 0 & 0  \\
  0 & 0 & 1 & 0 & 0 & 0 \\
\end{pmatrix}.
\ee
Written in the basis $\{L^{(0,\pm)},M^{(0,\pm)}\}$ the helicity operator simply assumes the form  $J_{a}P^{a}= \frac{1}{2}\left(L^{(+)} M^{(-)} + L^{(-)} M^{(+)} -2L^{(0)} M^{(0)}\right)$ and the corresponding nondegenerate $2$-form is
\be
\ka^{AB} = \frac{1}{4}  \begin{pmatrix}
  0 & 0 &  0 &  0 & 0 & 1  \\
  0 & 0 &   0 & 0 & -2& 0  \\
  0 & 0 & 0 & 1 & 0 & 0 \\
   0 &  0 &  1 &  0  & 0 & 0 \\
   0 &  -2 & 0 & 0 & 0 & 0  \\
   1 & 0 & 0 & 0 & 0 & 0 \\
\end{pmatrix}
\ee
which, when inverted, at once yields the quadratic form we are seeking
\be \label{CKng}
\ka_{AB} = 2 \begin{pmatrix}
  0 & 0 & 0 & 0 & 0 & 2  \\
  0 & 0 & 0 & 0 & -1& 0  \\
  0 & 0 & 0 & 2 & 0 & 0 \\
  0 & 0 & 2 & 0 & 0 & 0 \\
  0 & -1 & 0 & 0 & 0 & 0  \\
  2 & 0 & 0 & 0 & 0 & 0 \\
\end{pmatrix} .
\ee
The most general quadratic form defined on $\gca$ is then of the form
\be \label{CKgeneral}
\Om_{AB}=k \om_{AB}- \ka_{AB}=
 \begin{pmatrix}
  0 & 0 &  -2k &  0 & 0 & -2  \\
  0 & k &   0 & 0 & 1& 0  \\
  -2k & 0 & 0 & -2 & 0 & 0 \\
  0 &  0 &  -2 &  0  & 0 & 0 \\
  0 &  1 & 0 & 0 & 0 & 0  \\
  -2 & 0 & 0 & 0 & 0 & 0 \\
\end{pmatrix} 
\ee
where we have absorbed the factor of $2$ in Eq.(\ref{CKng}) into the arbitrary constant $k$ while taking the linear combination. It is straightforward to verify that the above bilinear form indeed satisfies the constraints in Eq.(\ref{cons}).  Now to explicitly evaluate the action given in Eq.(\ref{WZWact}) one requires to find the gauge fields $A^{A}_{\a}$'s. At this point, for definiteness, we need to fix a particular parametrization of the group element $g$ which we do as
\be 
g= e^{a_{+}L^{(+)}} e^{a_{0}L^{(0)}} e^{a_{-}L^{(-)}} e^{b_{+}M^{(+)}} e^{b_{0}M^{(0)}} e^{b_{-}M^{(-)}}.
\ee
By repeated use of the identity,
\be  \nn
e^{-A u} B e^{A u} = B-u[A,B] + \frac{u^2}{2!}[A,[A,B]] - \cdots
\ee
we find
\begin{align}
g^{-1} \partial_{\alpha}g &= L^{(-)}[a_{-}^2 e^{a_{0}} \partial_{\alpha}a_{+} + a_{-} \partial_{\alpha}a_{0} + \partial_{\alpha}a_{-} ] \nonumber \\ &
 + L^{(0)} [2a_{-} e^{a_{0}} \partial_{\alpha}a_{+} +
\partial_{\alpha}a_{0}]+ L^{(+)} [e^{a_{0}} \partial_{\alpha}a_{+}] \nonumber \\ & 
+ M^{(-)}[e^{a_{0}}\partial_{\alpha}a_{+} (2a_{-}b_{-}-a_{-}^2b_{0}) + \partial_{\alpha}a_{0} (b_{-}-b_{0}a_{-})-b_{0} \partial_{\alpha}a_{-}
+ \partial_{\alpha}b_{-} ] \nonumber\\& 
+ M^{(0)} [e^{a_{0}} \partial_{\alpha}a_{+}(2b_{-}-2a_{-}^2 b_{+})-2b_{+}a_{-} \partial_{\alpha}a_{0} -2b_{+} \partial_{\alpha}a_{-} + \partial_{\alpha}b_{0}] \nonumber\\& +
M^{(+)}[e^{a_{0}} \partial_{\alpha}a_{+} (b_{0}-2a_{-}b_{+})-b_{+} \partial_{\alpha}a_{0} + \partial_{\alpha}b_{+}]
\end{align}
from where we can easily extract out the gauge fields $A^{A}_{\a}$'s (making the identification given in Eq.(\ref{identification})), 
\bea \label{gaugefields}
\begin{aligned}
&A^{1}_{\a}= a_{-}^2 e^{a_{0}} \partial_{\alpha}a_{+} + a_{-} \partial_{\alpha}a_{0} + \partial_{\alpha}a_{-},\\
&A^{2}_{\a}=2a_{-} e^{a_{0}} \partial_{\alpha}a_{+} + \partial_{\alpha}a_{0},\\
&A^{3}_{\a}= e^{a_{0}} \partial_{\alpha}a_{+},\\
&A^{4}_{\a}= e^{a_{0}}\partial_{\alpha}a_{+} (2a_{-}b_{-}-a_{-}^2b_{0}) + \partial_{\alpha}a_{0} (b_{-}-b_{0}a_{-})-b_{0} \partial_{\alpha}a_{-}
+ \partial_{\alpha}b_{-}, \\
&A^{5}_{\a}= e^{a_{0}} \partial_{\alpha}a_{+}(2b_{-}-2a_{-}^2 b_{+})-2b_{+}a_{-} \partial_{\alpha}a_{0} -2b_{+} \partial_{\alpha}a_{-} + \partial_{\alpha}b_{0},\\
&A^{6}_{\a}= e^{a_{0}} \partial_{\alpha}a_{+} (b_{0}-2a_{-}b_{+})-b_{+} \partial_{\alpha}a_{0} + \partial_{\alpha}b_{+}.
\end{aligned}
\eea
Using Eqs.(\ref{CKgeneral},\ref{gaugefields}) one can then explicitly compute the terms in the action $S_{WZW}(g)$.
After some tedious calculation the final form of the action thus obtained is
\begin{align}
S_{WZW}(g) &= \frac{1}{4\pi}\int \limits_{\Si}d^{2}\s [\{-4ke^{a_{0}}\pa_{\a}{a_{-}}\pa_{\a}{a_{+}} -4\pa_{\a}a_{-}\pa_{\a}b_{+} +k(\pa_{\a}a_{0})^2 +2\pa_{\a}a_{0}\pa_{\a}b_{0}\nonumber\\
& -4a_{-}\pa_{\a}a_{0}\pa_{\a}b_{+} -4 e^{a_{0}}\pa_{\a}a_{+}\pa_{\a}b_{-} +4 a_{-}e^{a_{0}}\pa_{\a}a_{+}\pa_{\a}b_{0} -4  a_{-}^2e^{a_{0}} \pa_{\a}a_{+}\pa_{\a}b_{+})\}\nonumber\\
& +i\e_{\a \b}\{4ka_{-}{e^{a_{0}}}\pa^{\a}a_{0}\pa^{\b}a_{+} -4{a_{+}}{e^{a_{0}}}\pa^{\a}a_{0}\pa^{\b}b_{-} +4a_{-}\pa^{\a}a_{0}\pa^{\b}b_{+}\nonumber\\
& -4a_{-}{e^{a_{0}}}\pa^{\a}a_{+}\pa^{\b}b_{0}+4a^{2}_{-1}{e^{a_{0}}}\pa^{\a}a_{+}\pa^{\b}b_{+} \}].
\end{align}
We are now in a position to compare the above action with the $\s$-model action which is of the form
\be
S= \int d^2 \s \left(G_{MN} \pa_{\a}X^M \pa^{\a}X^N + iB_{MN} \e_{\a \b}\pa^{\a}X^M \pa^{\b}X^N \right)
\ee
where $X^M=\left\{a_{-},a_{0},a_{+},b_{-},b_{0},b_{+}\right\}$ are the coordinates on a $6$-dimensional manifold equipped with the metric $G_{MN}$ and $B_{MN}$ is the antisymmetric field.  Upon comparison we find that the WZW model describes a $6$-dimensional spacetime with the metric (omitting an overall factor of $1/4\pi$)
\begin{align}
ds^2 &= -4ke^{a_{0}}d{a_{-}}d{a_{+}} -4 da_{-}db_{+}
+k (da_{0})^2 +2 da_{0}db_{0} -4 a_{-}da_{0}db_{+} \nonumber\\&-4 e^{a_{0}}da_{+}db_{-} +4 a_{-}e^{a_{0}}da_{+}db_{0} -4  a_{-}^2e^{a_{0}} da_{+}db_{+}
\end{align}
which can be recast more compactly in matrix form as
\be
G_{MN} =  \begin{pmatrix}
  0 & 0 &  -2k{e^{a_{0}}} &  0 & 0 & -2  \\
  0 & k &   0 & 0 & 1& -2a_{-}  \\
  -2ke^{a_{0}} & 0 & 0 & -2e^{a_{0}} & 2a_{-}e^{a_{0}} & -2{a_{-}^2} e^{a_{0}} \\
  0 &  0 &  -2e^{a_{0}} &  0  & 0 & 0 \\
  0 &  1 & 2a_{-}e^{a_{0}} & 0 & 0 & 0  \\
  -2 & -2a_{-} & -2{a_{-}^2}e^{a_{0}} & 0 & 0 & 0 \\
\end{pmatrix}.
\ee
In a similar fashion, we can also read off the antisymmetric field tensor $B_{MN}$ which is found to have the following nonvanishing components
\bea
\begin{aligned}
&B_{23}= 2 k a_{-}{e^{a_{0}}}, \\
&B_{24}= - 2 {a_{+}}{e^{a_{0}}},\\
&B_{26}= 2 a_{-}, \\
&B_{35}= -2 a_{-}{e^{a_{0}}}, \\
&B_{36}=   2 {a_{-}^2}{e^{a_{0}}}.
\end{aligned}
\eea
Having found out the background spacetime metric and the antisymmetric fields let us turn to the conformal invariance of the theory. Being an WZW model, we expect our action to be conformally invariant. The $1$-loop $\b$-functions then must vanish (it can be shown that there is no contribution from higher order diagrams, see \cite{NW}),
\bea
\begin{aligned}
& \b^{G}_{MN}= R_{MN}-\frac{1}{4}H^{2}_{MN}-D_{M}D_{N}\phi=0,\\
&\b^{B}_{MN}= D^{L}H_{LMN} +D^{L}\phi H_{LMN} =0, \\
& \b^{\phi}= -R + \frac{1}{12} H^2 + 2 \nabla^2 \phi + (\nabla \phi)^2 + \Lambda = 0.
\end{aligned}
\eea
where $\phi$ is the dilaton and we have defined, $H_{LMN}= D_{[L}B_{MN]},H^{2}_{MN} = H_{MPR}H^{PR}_N, H^2= H_{MNR}H^{MNR}$ and $\La=2(c-6)/3\a'$ with $c$ being the central charge and $\sqrt{\a'}=l_{s}$, is the string length. The nonvanishing components of $R_{MN}$ are $R_{13}= 2{e^{a_{0}}}$ and $R_{22}= -1$. The Ricci scalar $R$ vanishes and so does $H^2$. The surviving components of $H^{2}_{MN}$ are $H^{2}_{13}=8 e^{a_{0}}$ and $H^{2}_{22}=-4$. Collecting all these together one can easily verify that the $\b$-functions indeed vanish with $\phi = \text{constant}$ and $c=6$. This is in conformity with earlier results that the Virasoro central charge is an integer equal to the dimension of the group manifold \cite{Sfe,OliRab,Sfe2,KehMee,FigSta}. For WZW models, it is also possible to show the conformal invariance in a nonperturbative way by carrying out the Sugawara construction. However, the model lends itself to such a construction only if there exists a nondegenerate bilinear form. Since, in the present case the algebra is nonsemisimple and the Killing-Cartan form becomes noninvertible the standard Sugawara construction is no longer feasible.
However, we can obtain the following invariant form
\be 
L^{AB}=  \begin{pmatrix}
  0 & 0 &  0 &  0 & 0 & -\frac{1}{4}  \\
  0 & 0 &   0 & 0 & \frac{1}{2}& 0  \\
  0 & 0 & 0 & -\frac{1}{4} & 0 & 0 \\
   0 &  0 &  -\frac{1}{4} &  0  & 0 & \frac{k}{4} + \frac{1}{8} \\
   0 & \frac{1}{2}  & 0 & 0 & -\frac{k}{2}-\frac{1}{4} & 0  \\
   -\frac{1}{4} & 0 & 0 & \frac{k}{4} + \frac{1}{8} & 0 & 0 \\
\end{pmatrix} 
\ee
obtained as a solution to (see \cite{KehMee})
\be 
L^{AB}\left(2\Om_{BC}+ \om_{BC} \right)=\de^{A}_{~C}
\ee
which 
satisfies the Virasoro master equation \cite{HalKir}
\be
L^{AB}=2L^{AC} \Omega_{CD} L^{DB}-L^{CD}L^{EF} {f^A}_{CE} {f^B}_{DF}-L^{CD} {f^F}_{CE} {f^{[A}}_{DF} L^{B]E}
\ee
thereby bearing out the conformal invariance of the theory. The central charge is
\be 
c=2L^{AB}\Om_{AB}=6
\ee
as anticipated.\\
Finally, let us get back to a point we had raised at the beginning - that WZW models provide exact string backgrounds. While the background we have obtained is an exact string background, it is plagued by the presence of three timelike directions, as evident from the signature of the metric, which is $(+++---)$.
This is, of course, not a very physically appealing situation. In fact, there are earlier instances (see, for example, \cite{Sfe} where $d$-dimensional centrally extended Euclidean algebra was considered, or \cite{Keh}) where such a metric with more than one timelike direction has been encountered. In such cases one usually gauges an appropriate subgroup \cite{Keh} to eliminate the unwanted timelike directions and arrive at a physically meaningful spacetime. It will be interesting to see whether the same can be done in the present case so that we can do away with two timelike directions and arrive at a spacetime with a signature $(+++-)$. If this is indeed possible the resulting background might describe a more realistic string theory in $(3+1)$-spacetime dimension. We leave this issue for future consideration.\\
Before closing this note, let us also comment on a potentially very interesting avenue of future research. Our construction of the non-degenerate $2$-form above relied on the isomorphism between $\gca$ and the Poincar\'e algebra in $(2+1)$-dimension, $iso(2,1)$. This isomorphism actually extends to the infinite dimensional $2$-dimensional GCA (mentioned previously) and the $3$-dimensional Bondi-Metzner-Sachs algebra which is the asymptotic symmetry algebra of $3$-dimensional Minkowski spacetime \cite{BagPRL}. This isomorphism has been recently exploited in an attempt to understand flat-space holography \cite{BagFar, BagDet, BagDet2}. With our construction of the WZW model of $\gca$ in this paper, we would like to further this program. It is well-known that a Chern-Simons theory on a manifold with a boundary induces a dynamical chiral WZW theory on the boundary \cite{Moore, Elit}.  Gravity in $3$-dimension can be expressed as a Chern-Simons theory. Hence our construction of a WZW model of $\gca$ could be used in connection with a Chern-Simons formulation of flat-space gravity in $3$-dimension. Carlip \cite{Car} has used the Chern-Simons/WZW connection in $AdS_{3}$ to account for the entropy of BTZ black holes by a counting of states in the WZW theory. Analogues of the BTZ black holes have been recently identified in $3$-dimensional flat space as cosmological solutions with horizons \cite{BagDet2, Bar}. The entropy of these objects have been reproduced by a Cardy-like counting in the dual field theory \cite{BagDet2} (see also \cite{Bar2}). It would be very interesting to evoke Carlip's original arguments in this context to re-derive the entropy of these "flat-BTZ"s in terms of a counting of states in the WZW model of $\gca$.

\vspace{.2cm}

\section*{Acknowledgements}

The authors would like to thank Arjun Bagchi who has been the primary driving force behind this project and Shibaji Roy for various fruitful discussions. SC would like to thank Palash Baran Pal for valuable discussions. PD would like to acknowledge thankfully  the financial support of the Council of Scientific and Industrial Research, India (SPM-07/489 (0089)/2010-EMR-I).


\vspace{.5cm}

\begin{thebibliography}{99}
\bibitem{NW}
C. R. Nappi, E. Witten,
Phys. Rev.  Lett.\ {\bf 71}, 23 (1993).
\bibitem{Sfe}
K. Sfetsos,
[hep-th/9311093].
\bibitem{OliRab}
D.I. Olive, E. Rabinovici, A. Schwimmer,
Phys. Lett. {\bf B321}, 361 (1994).
\bibitem{Sfe1}
K. Sfetsos,
Phys. Rev. {\bf D50}, No. 4, 2784 (1994).
\bibitem{Sfe2}
K. Sfetsos,
Phys. Lett. {\bf B324}, 335 (1994).
\bibitem{Moh}
N. Mohammedi,
Phys. Lett. {\bf B325}, 371 (1994).
\bibitem{KehMee}
A.A. Kehagias, P.A.A. Meessen,
Phys. Lett. {\bf B331}, 77 (1994).
\bibitem{Keh}
A.A. Kehagias,
[hep-th/9406136].
\bibitem{Baru}
A.O. Barut,
Helv. Phys. Acta {\bf 46}, 496 (1973).
\bibitem{HavPleb}
P. Havas, J. Pleba\'nski,
 J.  Math. Phys. {\bf 19}, 482 (1978).
\bibitem{HenUnt}
M. Henkel, J. Unterberger,
Nucl. Phys. {\bf B660}, 407 (2003).
\bibitem{LukStich}
J. Lukierski, P.C. Stichel, W.J. Zakrzewski,
Phys. Lett. {\bf A357}, 1 (2006).
\bibitem{DuvHor}
C .Duval, P.A. Horvathy,
J. Phys. {\bf  A42}, 465206 (2009).
\bibitem{DuvHor2}
C. Duval, P. Horvathy,
 J. Phys. {\bf A44}, 335203 (2011).
\bibitem{BagGop}
A. Bagchi, R. Gopakumar,
JHEP {\bf 07}, 037 (2009).
\bibitem{BagGop1}
A. Bagchi, R. Gopakumar, I. Mandal, A. Miwa,
JHEP {\bf 08}, 004 (2010).
\bibitem{BagKun}
A. Bagchi, A. Kundu,
Phys. Rev. {\bf D83}, 066018 (2011).
\bibitem{FigSta}
J.M. Figueroa-O'Farrill, S. Stanciu,
Phys. Lett. {\bf B327}, 40 (1994).
\bibitem{BagPRL}
A. Bagchi,
Phys. Rev. Lett. {\bf 105}, 171601 (2010).
\bibitem{HalKir}
M.B. Halpern, E. Kiritsis,
Mod. Phys. Lett. {\bf A4}, 1373 (1989).
\bibitem{BagFar}
A. Bagchi, R. Fareghbal,
[hep-th/1203.5795].
\bibitem{BagDet}
A. Bagchi, S. Detournay, D. Grumiller,
[hep-th/1208.1658].
\bibitem{BagDet2}
A. Bagchi, S. Detournay, R. Fareghbal, J. Simon,
[hep-th/1208.4372].
\bibitem{Moore}
G. Moore, N. Seiberg,
Phys. Lett.  {\bf B220}  422 (1989).
\bibitem{Elit}
S. Elitzur \textit{et al.},
Nucl.  Phys. {\bf B326} 108 (1989).
\bibitem{Car}
S. Carlip,
Phys. Rev. {\bf D51} 632 (1995).
\bibitem{Bar}
G. Barnich, A. Gomberoff, H.A. Gonz\'alez,
[hep-th/1204.3288].
\bibitem{Bar2}
G. Barnich,
[hep-th/1208.4371].
\end{thebibliography}
\end{document}